\documentclass[conference]{IEEEtran}
\IEEEoverridecommandlockouts
\usepackage{cite}
\usepackage{amsmath,amssymb,amsfonts}
\usepackage{algorithmic}
\usepackage{graphicx}
\usepackage{textcomp}
\usepackage{xcolor}
\usepackage{listings}
\usepackage{url}
\usepackage{booktabs}
\usepackage{comment}
\usepackage{subfig}

\def\BibTeX{{\rm B\kern-.05em{\sc i\kern-.025em b}\kern-.08em
    T\kern-.1667em\lower.7ex\hbox{E}\kern-.125emX}}

\begin{document}

\title{From Sensors to Insight: Rapid, Edge-to-Core Application Development for Sensor-Driven Applications}

\author{\IEEEauthorblockN{Komal Thareja\textsuperscript{*}, Anirban Mandal\textsuperscript{*}, Ewa Deelman\textsuperscript{\ddag}}
\IEEEauthorblockA{\textsuperscript{*}Renaissance Computing Institute, University of North Carolina at Chapel Hill, NC, USA}
\IEEEauthorblockA{\textsuperscript{\ddag}Information Sciences Institute, University of Southern California, Marina del Rey, CA, USA}
}

\maketitle
\begin{abstract}
    
Scientists increasingly rely on sensor-based data, yet transforming raw streams into insights across the edge-to-cloud continuum remains difficult. Provisioning heterogeneous infrastructure and managing execution on emerging platforms like Data Processing Units typically requires cross-domain expertise, creating significant barriers to rapid prototyping.

This paper introduces an experience-driven methodology for the rapid development of sensor-driven applications. By combining pattern-based workflow engineering with AI-assisted development—implemented via Pegasus on the FABRIC testbed—we utilize an existing Orcasound hydrophone workflow as a reusable template. We introduce a pattern-based engineering methodology to generate and refine workflows for air quality, earthquake, and soil moisture monitoring. Furthermore, we show how these abstract structures are extended to edge resources through modular configuration and placement.
Our evaluation focuses on user productivity and practical lessons rather than peak performance. Through these case studies, we illustrate how AI-assisted, pattern-based development lowers the entry barrier for non-experts and enables iterative exploration of sensor-driven applications across distributed infrastructures.
\end{abstract}


\section{Introduction}

Sensor-driven applications are a cornerstone of cutting-edge research in environmental monitoring, smart infrastructure, marine biology, and cyber-physical systems. These applications increasingly demand end-to-end processing pipelines that transform raw sensor data into actionable insight through a combination of edge processing, distributed analytics, and cloud-scale computation. Although sensing hardware, networking, and analytics frameworks have advanced rapidly, the development of integrated edge-to-core applications remains complex, error-prone, and time-consuming.

A key challenge lies in bridging the gap between sensing systems and scalable computation---what we term the \textbf{``blank page bottleneck.''}  Traditionally, moving from a scientific question (e.g., ``How does local air quality impact respiratory health trends?'') to a production scientific workflow capable of generating data to answer this questions, requires deep infrastructure expertise. Instead of exploring their data, scientists often find themselves bogged down with execution mechanics like workflow formats, container orchestration, and infrastructure provisioning. For many practitioners, particularly domain scientists and IoT application developers, this breadth of required knowledge presents a significant barrier to rapid prototyping and deployment.

Workflow management systems such as Pegasus~\cite{deelman2015pegasus} and programmable infrastructures like FABRIC~\cite{baldin2019fabric} offer powerful abstractions for managing complex distributed applications, including emerging hardware accelerators such as Data Processing Units (DPUs)~\cite{bluefield2022}. However, effectively leveraging these tools requires the understanding of configuration models, deployment options, and infrastructure interfaces---specifically for sensor-driven or edge-centric applications~\cite{thain2005distributed}.

We address this challenge through \textbf{pattern-based engineering with AI assistance}. Rather than starting from scratch, users provide an AI assistant (e.g. Claude~\cite{anthropic_claude}) with an existing Pegasus workflow pattern and describe their new use case. The AI adapts the pattern---retaining dependency logic while swapping domain-specific components---producing a working first draft. This shifts development from \textit{code-first} to \textit{intent-first}: scientists describe their data sources, analysis steps, and expected outputs in plain English, and AI generates production-ready workflow code~\cite{chen2021evaluating}. Grounding AI in existing patterns prevents hallucinations and ensures that the generated code follows established best practices.

We operationalize this approach through a \textbf{5-step development loop} that compresses days of workflow creation to minutes: (1) \textit{Describe Science}---define data, analysis steps, and outputs in plain English; (2) \textit{Ground with Context}---provide the AI assistant with a proven Pegasus example; (3) \textit{Generate Skeleton}---AI produces working Python code with task definitions, I/O declarations, and dependency graphs; (4) \textit{Refine}---human-in-the-loop iteration to adjust resource requirements and parallelization; (5) \textit{Scale}---deploy to the edge, clusters or clouds without redesign.

We demonstrate this methodology through four representative scientific workflows spanning diverse sensor domains and execution patterns: \textbf{air quality forecasting} using OpenAQ~\cite{openaq} data with LSTM models, \textbf{earthquake analysis} with USGS~\cite{usgs} seismic data and branching analytical pipelines, \textbf{Orcasound} hydrophone~\cite{orcasound2020} processing for Orca whale detection with edge-to-cloud computation including DPU offload, and \textbf{soil moisture prediction}~\cite{soilmoisture-workflow} for precision agriculture. These workflows are deployed on FABRIC infrastructure using FABlib~\cite{fablib2023} for provisioning, HTCondor~\cite{thain2005distributed} for job scheduling, and Pegasus for workflow orchestration.

This paper makes the following contributions:
\begin{itemize}
  \item We present a \textbf{pattern-based engineering methodology} for rapid edge-to-core application development, demonstrating how AI assistance combined with existing workflow tools (Pegasus/HTCondor) and infrastructure platforms (FABRIC/FABlib) accelerates development of sensor-driven scientific applications.
  \item We describe a \textbf{5-step development loop} that shifts workflow construction from code-first to intent-first, enabling scientists to describe analyses in plain English while AI generates production-ready Pegasus workflows.
  \item We provide practical insights from \textbf{four case studies} spanning air quality, seismology, marine bioacoustics, and agriculture---deployed across edge-to-core configurations including BlueField-3 DPUs---demonstrating flexibility in execution placement and reduced development effort.
  \item We articulate \textbf{why Pegasus and AI are synergistic}: the explicit, readable, composable, and execution-agnostic nature of Pegasus workflows makes them ideal targets for LLM-assisted generation and adaptation.
\end{itemize}

The remainder of this paper is organized as follows. Section~\ref{sec:background} presents background and related work on scientific workflows, programmable research infrastructure, edge-to-core computing, and AI-assisted development. Section~\ref{sec:methodology} describes our pattern-based, AI-assisted methodology for constructing sensor-driven workflows on programmable infrastructure. Section~\ref{sec:case-studies} presents a sequence of case studies demonstrating progressive workflow reuse, incremental development, and extension to edge resources. Section~\ref{sec:evaluation} provides a user-centric evaluation focusing on iteration effort, debugging experience, and time-to-first-execution. Section~\ref{sec:conclusion} presents our conclusions and directions for future work.

\section{Background and Related Work}
\label{sec:background}

%This work builds on prior research in scientific workflow management, programmable research infrastructure, edge-to-core computing, and AI-assisted software development. %Rather than advancing any single area in isolation, our contribution lies in combining these components into a practical, developer-centric methodology for building sensor-driven applications. 
%We briefly review relevant background and position our work with respect to existing efforts.
We briefly review relevant background and position our work with respect to prior research in scientific workflow management, programmable research infrastructure, edge-to-core computing, and AI-assisted software development.

%\subsection{Scientific Workflow Management Systems}
\vspace{5pt}
\noindent {\bf Scientific Workflow Management Systems.}
Scientific workflow management systems address the complexity of coordinating multi-stage computations across distributed resources. Systems such as Pegasus~\cite{deelman2015pegasus}, Makeflow~\cite{albrecht2012makeflow}, Kepler~\cite{ludascher2006kepler}, 
NextFlow~\cite{diTommaso2017nextflow}, and Airflow~\cite{airflow} allow users to express applications as workflows with explicit task dependencies, enabling %automation of 
automated data movement, job scheduling, and fault handling.

Pegasus is particularly well suited for distributed scientific applications due to its separation of abstract workflow descriptions from execution environments. Users define workflows as directed acyclic graphs (DAGs) independent of infrastructure, while Pegasus handles planning, data staging, retries, and provenance tracking at execution time. Pegasus commonly integrates with HTCondor~\cite{thain2005distributed}, which provides a flexible execution substrate capable of managing heterogeneous resources through declarative scheduling constraints.

Prior work has demonstrated the scalability, portability, and reliability benefits of workflow systems in high-performance and distributed computing environments ~\cite{deelman2019evolution, vahi2018workflows, do2022accelerating}. However, constructing and configuring workflows remains a significant barrier for new users, especially when workflows span multiple execution sites or involve heterogeneous resources such as edge devices or accelerators.

%\subsection{Programmable Research Infrastructure}
\vspace{5pt}
\noindent {\bf Programmable Research Infrastructure.}
Programmable research infrastructures provide researchers with fine-grained control over compute, storage, and networking resources. Platforms such as FABRIC~\cite{baldin2019fabric}, Chameleon~\cite{dupont2019chameleon}
, and CloudLab~\cite{dupont2019cloudlab}, 
enable users to deploy custom topologies, experiment with advanced networking features, and evaluate distributed systems under realistic conditions.

FABRIC extends these capabilities by offering a geographically distributed testbed with support for heterogeneous hardware, including GPUs, FPGAs, SmartNICs, and DPUs. Access is provided through programmable APIs and notebooks, allowing infrastructure to be provisioned and reconfigured as code. While such platforms enable powerful experimentation, effectively leveraging them for scientific applications often requires expertise in infrastructure configuration, networking, and system integration.

Our work operates within this context, using programmable infrastructure not as an end in itself, but as a substrate on which sensor-driven workflows can be deployed
%, adapted, 
and evaluated.

%\subsection{Edge-to-Core and IoT Computing}
\vspace{5pt}
\noindent {\bf Edge-to-Core Computing.}
Edge computing distributes computation closer to data sources to reduce latency, conserve bandwidth, and improve responsiveness~\cite{shi2016edge}. In sensor-driven and IoT applications, edge processing is often combined with cloud-based analytics, forming an edge-to-core continuum. Prior research has explored task placement, data offload strategies, and resource-aware scheduling across this continuum.

Recent hardware trends, including programmable SmartNICs~\cite{firestone2018smartnic} and DPUs~\cite{bluefield2022}, further blur the boundary between networking and computation by enabling data processing at the network edge. These platforms have been used to accelerate I/O-intensive tasks and offload preprocessing from host CPUs. However, integrating such resources into end-to-end sensor application workflows typically requires custom orchestration logic or specialized frameworks.

In contrast, our approach demonstrates how existing workflow systems can incorporate edge devices and DPUs using standard scheduling and placement mechanisms, avoiding bespoke edge orchestration layers.

%\subsection{AI-Assisted Software and Workflow Development}
\vspace{5pt}
\noindent {\bf AI-Assisted Software and Workflow Development.}
Large language models have recently been applied to software development tasks such as code generation, debugging, and API exploration. Tools including GitHub Copilot, ChatGPT, Claude, and Gemini have been shown to improve user productivity, particularly for boilerplate generation and error diagnosis~\cite{chen2021evaluating}. Several recent efforts explore the use of AI for infrastructure-as-code generation and workflow specification. However, many such approaches treat AI-generated scripts as standalone artifacts, limiting portability, provenance tracking, and fault tolerance. Moreover, systematic studies of AI-assisted development for scientific workflows—especially in edge-to-core sensing contexts—remain limited.

Our work differs in two key ways. First, AI assistance is explicitly grounded in existing, validated workflow patterns rather than unconstrained generation. Second, AI is used to assist workflow construction and refinement within mature workflow systems, preserving their benefits while reducing the learning curve for new users.

%\subsection{Positioning}

\vspace{5pt}
\noindent {\bf Positioning.} In summary, prior work has established the value of workflows, programmable infrastructure, edge computing, and AI-assisted development independently. However, there is limited empirical guidance on how these components can be combined to support \textit{rapid, iterative development of sensor-driven applications across the edge-to-core continuum}, particularly from the perspective of user productivity.
%and accessibility.

This paper addresses this gap by presenting an experience-driven methodology that integrates AI-assisted development with pattern-based workflow engineering on programmable infrastructure. Rather than proposing new workflow abstractions or scheduling algorithms, we focus on practical lessons learned from building and extending real applications, providing actionable insights for researchers and practitioners working at the intersection of sensing, workflows, and distributed systems.

\section{Methodology}
\label{sec:methodology}

\begin{figure}[!t]
  \centering
  \includegraphics[width=0.95\linewidth]{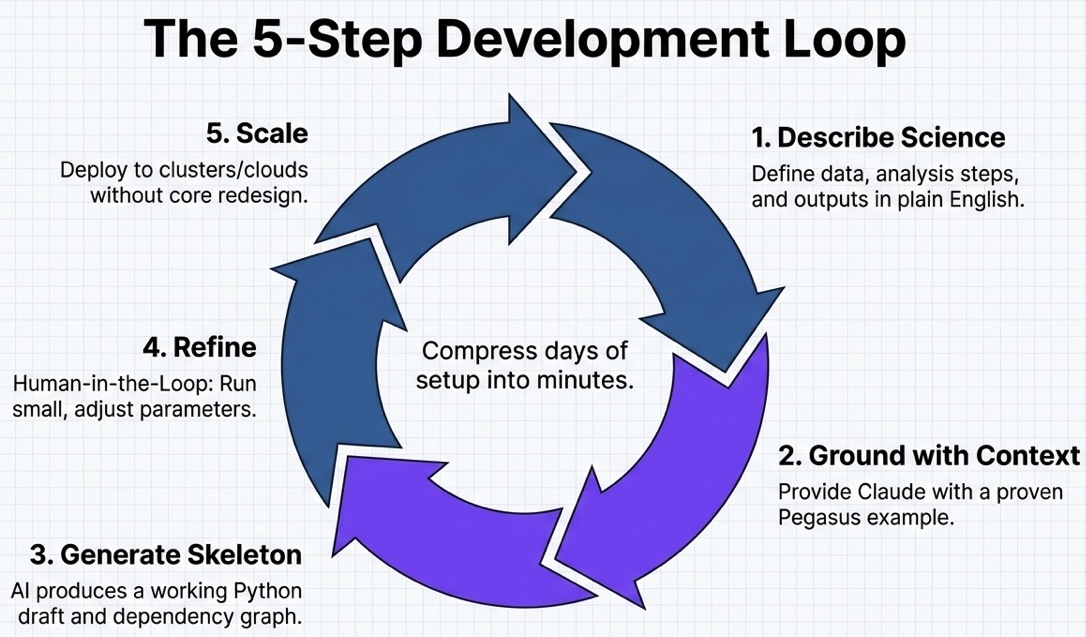}
  \caption{Pattern-based, AI-assisted development loop for sensor-driven edge-to-core applications.}
  \vspace{-2pt}
  \label{fig:five_step_loop}
\end{figure}

This section describes our methodology for rapidly developing sensor-driven, edge-to-core scientific applications by combining \textit{pattern-based workflow engineering} with \textit{AI-assisted development}, implemented using Pegasus, HTCondor, and programmable infrastructure on FABRIC. The methodology is derived from practical experience building and deploying real-world sensor workflows and is designed to reduce user effort while preserving the rigor, portability, and reproducibility of workflow-based execution. Figure~\ref{fig:five_step_loop} summarizes the five-step, AI-assisted development loop used to construct, refine, and deploy sensor workflows across edge-to-core resources.

%\begin{figure}[t]
%  \centering
%  \includegraphics[width=0.95\linewidth]{figures/five_step_rule.jpg}
%  \caption{Pattern-based, AI-assisted development loop for sensor-driven edge-to-core applications.}
%  \label{fig:five_step_loop}
  %\vspace{-10pt}
%\end{figure}

%\subsection{Design Principles}
Our methodology is guided by four {\it design principles}: \textbf{(1) Intent-first development:} Application development begins with a scientific intent---data sources, analysis steps, and expected outputs---expressed in natural language rather than low-level execution mechanics; \textbf{(2) Pattern reuse over greenfield design:} New applications are derived by adapting existing, validated workflow patterns rather than starting from an empty specification, mitigating the ``blank page bottleneck.''; \textbf{(3) Separation of concerns:} Workflow logic, execution placement, and infrastructure provisioning are explicitly separated. Abstract workflows remain execution-agnostic, while placement decisions are expressed in configuration and scheduling layers; \textbf{(4) Human-in-the-loop AI assistance:} AI tools assist with generation, adaptation, and debugging but do not replace workflow engines or scheduling systems; users retain control and validate changes through execution.

\begin{comment}
\textbf{Intent-first development:} Application development begins with a scientific or operational intent---data sources, analysis steps, and expected outputs---expressed in natural language rather than low-level execution mechanics.

\textbf{Pattern reuse over greenfield design:} New applications are derived by adapting existing, validated workflow patterns rather than starting from an empty specification, mitigating the ``blank page bottleneck.''

\textbf{Separation of concerns:} Workflow logic, execution placement, and infrastructure provisioning are explicitly separated. Abstract workflows remain execution-agnostic, while placement decisions are expressed in configuration and scheduling layers.

\textbf{Human-in-the-loop AI assistance:} AI tools assist with generation, adaptation, and debugging but do not replace workflow engines or scheduling systems; developers retain control and validate changes through execution.
\end{comment}

\subsection{The Five-Step Development Loop}
Our approach follows an iterative five-step loop that compresses workflow development from days to hours:
%of setup into rapid cycles:

\textbf{Step 1: Describe the science.}
Scientists describe the application intent in plain language, including sensor data sources (e.g., APIs or object storage), processing stages (preprocessing, feature extraction, inference), and outputs (forecasts, alerts, visualizations). Infrastructure specifics are intentionally omitted.

\textbf{Step 2: Ground with an existing workflow pattern.}
Instead of generating from scratch, users provide a reference Pegasus workflow that captures a similar structure (e.g., ingest $\rightarrow$ process $\rightarrow$ predict $\rightarrow$ visualize; branching analyses; or edge-to-cloud splits). The AI assistant is instructed to preserve the dependency structure while adapting task semantics, data access, and executables.

\textbf{Step 3: Generate a production-ready workflow skeleton.}
Given the grounded pattern, the AI assistant produces executable Pegasus Python code~\cite{pegasus_api_reference} containing: (i) task definitions, (ii) explicit input/output declarations (Replica and Transformation catalogs), (iii) dependency relationships (DAG topology), and (iv) parameterized loops over sensors, regions, or time windows.

\textbf{Step 4: Human-guided refinement.}
Scientists iteratively refine resource sizing (cores/memory), containerization, and parallelization granularity. AI assistance is used to interpret Pegasus planning failures, HTCondor logs, and configuration issues, but final edits are reviewed and validated by the user.

\textbf{Step 5: Scale and re-target execution.}
After local validation, the same abstract workflow is scaled or re-targeted to different execution environments without redesign. Execution placement is controlled through Pegasus catalogs and properties, enabling the same workflow to run across laptops, clusters, and heterogeneous edge-to-core deployments.

\subsection{Workflow Construction and Execution Model}
All applications are represented as Pegasus DAGs with explicit data dependencies. This representation is particularly compatible with AI-assisted development because it is \textit{explicit}, \textit{readable}, \textit{composable}, and \textit{execution-agnostic}. Pegasus plans the abstract workflow onto available resources, while HTCondor performs job dispatch and placement.

For edge-to-cloud workflows, HTCondor requirements (ClassAds) steer tasks to appropriate nodes (e.g., DPU-enabled edge workers vs.\ CPU-only cloud workers)~\cite{pegasus_edge}. The ClassAds are written automatically by Pegasus. Containerized codes further decouple workflow logic from runtime environments, improving reproducibility and portability across heterogeneous sites.

\subsection{Infrastructure Provisioning for Edge-to-Core Deployments}
Infrastructure is provisioned programmatically using FABlib. A typical deployment includes: (i) a submit node hosting Pegasus and the HTCondor central manager, (ii) edge workers near data sources (optionally equipped with DPUs), and (iii) cloud/cluster workers for compute-intensive stages. Connectivity is established using FABRIC networking services, enabling reproducible and reconfigurable experimental environments.
In workflows incorporating DPUs, I/O-intensive preprocessing (e.g., audio format conversion, feature extraction) is placed on DPU-attached ARM cores, while ML inference and aggregation execute on general-purpose compute nodes.

\subsection{Role of AI Assistance}
%AI assistance is applied \textit{exclusively at development time} to accelerate: workflow adaptation from patterns, API exploration and boilerplate generation, debugging of planning/execution errors, and incremental workflow extension. The AI assistant does not execute workflows or make scheduling decisions; 
AI assistance is applied \textit{exclusively at development time} to accelerate workflow adaptation, API exploration, boilerplate generation, and debugging—not workflow execution or scheduling. It serves as a productivity layer that lowers the barrier to using Pegasus/HTCondor and FABRIC while preserving provenance, fault tolerance, and portability. All prompts are documented in the repository's \texttt{README.md} for reproducibility.%The prompts used to interact with the AI models are documented in the workflow repository’s \texttt{README.md} for transparency and reproducibility.

\section{Case Studies: Pattern Reuse and AI-Assisted Workflow Evolution}
\label{sec:case-studies}

We present our case studies as a \textit{progressive development narrative} rather than isolated application examples. Consistent with our goals, the purpose of these case studies is not to evaluate peak application performance, but to examine how effectively our methodology supports \textit{rapid workflow construction, incremental refinement, and extension across domains and execution environments}. We emphasize user productivity (time-to-first-workflow, iteration effort, and portability of the workflow pattern) and highlight where AI assistance reduced friction during development.

Across all case studies, Pegasus workflows were developed and executed on a Pegasus/HTCondor environment deployed on the FABRIC testbed. We additionally extended selected workflow stages to edge resources, including FABRIC DPUs and Raspberry Pis, to demonstrate how edge devices can participate in end-to-end workflows without redesigning the workflow logic.

\subsection{Baseline Workflow Pattern: Orcasound as Starting Point}
\label{subsec:orcasound-pattern}

The \textbf{Orcasound hydrophone processing workflow} served as the foundational pattern for subsequent workflow development. Importantly, this workflow was already available and operational prior to this effort, and we used it as the reference implementation to ground AI-assisted workflow generation.

Orcasound represents a canonical sensor-driven analytics structure common across multiple domains~\cite{orcasound2020}:
\[
\textit{ingest} \rightarrow \textit{preprocess} \rightarrow \textit{transform} \rightarrow \textit{infer} \rightarrow \textit{aggregate}.
\]
Concretely, this sensor-driven workflow ingests audio segments, 
%from object storage, 
converts them into analysis-ready formats, performs ML inference to detect Orca vocalizations, and aggregates predictions across sensors and time windows. We treat Orcasound not only as an application but as a reusable \textit{workflow pattern} representative of many sensor-driven applications: its explicit dependency graph, containerized transformations, and separation of workflow logic from execution placement make it well suited for adaptation across other domains.

\subsection{Pattern Reuse Across Domains with AI Assistance}
\label{subsec:pattern-reuse}

Using Orcasound~\cite{orcasound-workflow} as a reference pattern, we generated three new sensor workflows---\textbf{air quality analysis}~\cite{airquality-workflow}, \textbf{earthquake analysis}~\cite{earthquake-workflow}, and \textbf{soil moisture prediction}~\cite{soilmoisture-workflow}---by preserving the high-level dependency structure while adapting domain-specific tasks, data sources, and parameters.

\begin{figure*}[t]
  \centering
  \subfloat[Air quality workflow: ingestion, feature engineering, modeling, delivery.]{
    \includegraphics[width=0.48\textwidth]{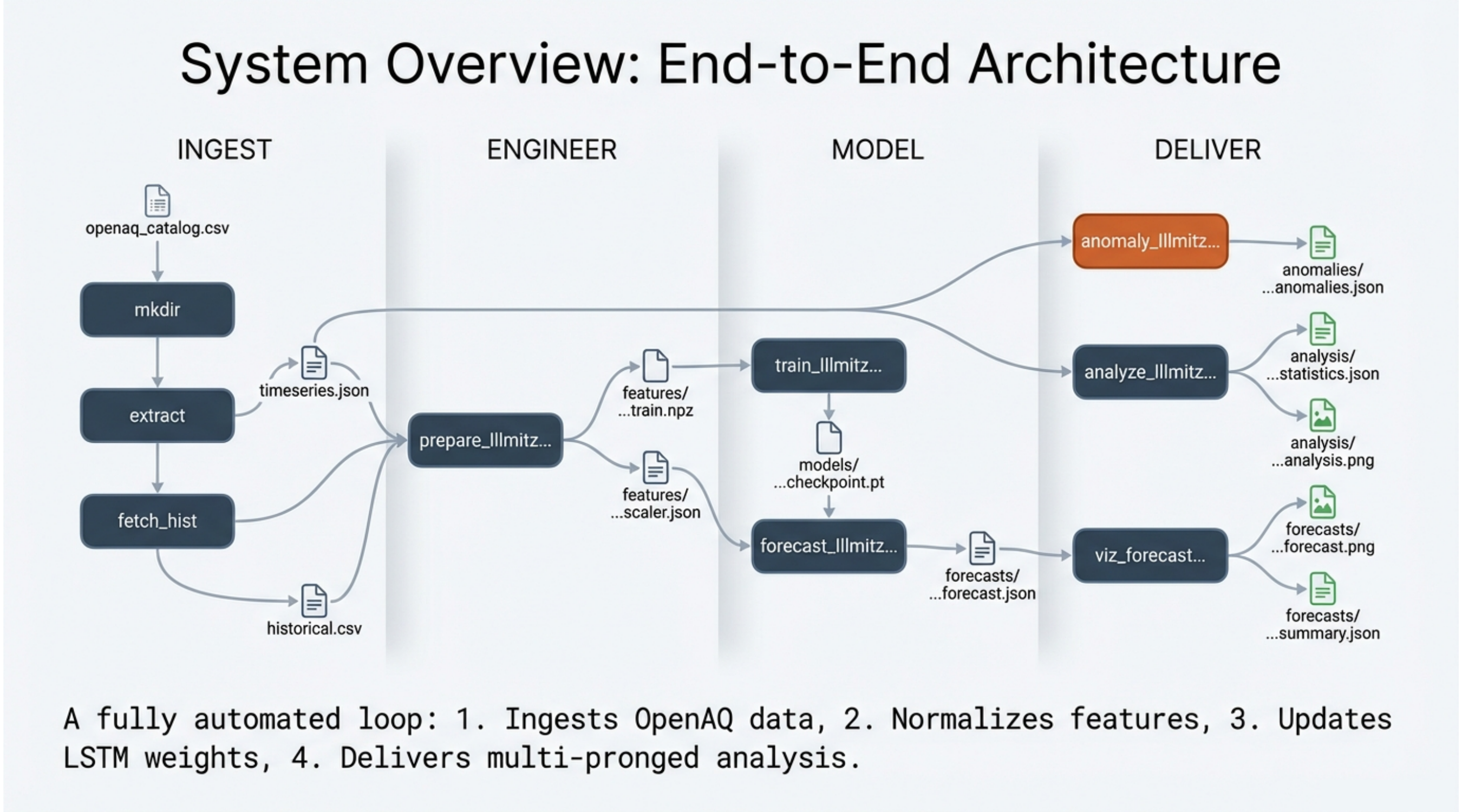}
    \label{fig:airquality-dag}
  }
  \hfill
  \subfloat[Representative air quality forecast and anomaly detection output.]{
    \includegraphics[width=0.48\textwidth]{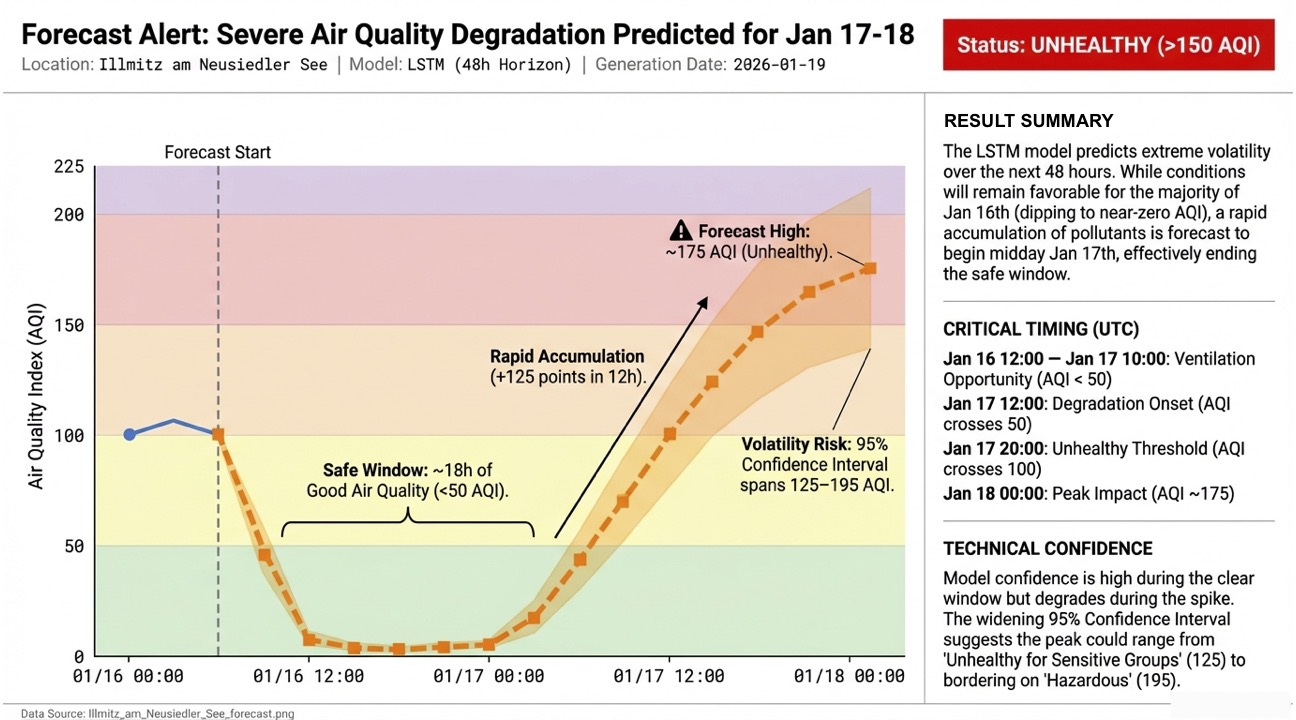}
    \label{fig:airquality-results}
  }
  \vspace{-8pt}
  \caption{Air quality workflow structure and representative analytical output.}
  \label{fig:airquality-workflow-results}
  \vspace{-20pt}
\end{figure*}

Rather than manually designing new workflows from scratch, we followed an incremental prompting strategy:
(i) provide the AI assistant like Claude (Opus 4.5 Model) with the Orcasound Pegasus workflow as a concrete template,
(ii) describe the new scientific goal and data source in natural language, and
(iii) instruct the assistant to retain the workflow structure while swapping domain components (ingestion logic, preprocessing steps, model/inference stages, and outputs).
This grounded approach enabled rapid creation of runnable Pegasus workflow skeletons and reduced trial-and-error in configuration and dependency specification.

\subsection{Air Quality Workflow: Incremental Development Trajectory}
\label{subsec:airquality-trajectory}

The air quality workflow illustrates how AI-assisted development supports stepwise workflow evolution from a simple, working pipeline to a richer ML-based workflow, without destabilizing earlier stages. Figure~\ref{fig:airquality-dag} shows the resulting workflow structure after incremental extensions, highlighting how new stages were added without restructuring the existing dependency graph.

\subsubsection{Initial baseline pipeline (ingest--analyze--detect anomaly)}

The starting point for this workflow was an existing Orcasound acoustic monitoring template~\cite{orcasound2020}, which had previously been expressed as a Pegasus workflow capturing a simple ingest–process–detect pattern. Rather than designing the air quality workflow from scratch, we provided this template to the AI assistant as a structural reference and expressed a new scientific intent: monitoring particulate matter levels in air quality data instead of detecting anomalous acoustic signals. The assistant mapped this intent onto the existing workflow skeleton by identifying analogous stages—data ingestion, signal analysis, and anomaly detection—and adapting them to the new domain. 

In this phase, the input data were sourced solely from OpenAQ. The AI assistant generated a Pegasus workflow that (i) fetched particulate matter measurements for a specified region and time window, (ii) performed basic aggregation/cleaning, and (iii) detected anomalous readings using lightweight statistical rules (e.g., thresholds or rolling statistics). This baseline provided a fast \textit{time-to-first-execution} milestone and created a stable foundation for later extensions.

\subsubsection{Extending to ML-based forecasting}
After validating the baseline pipeline, we prompted the AI assistant to extend the workflow with ML-based forecasting. The assistant introduced additional stages for feature preparation, model training (or model update), and prediction while preserving upstream ingestion and preprocessing logic. Crucially, this extension required no redesign of the existing pipeline structure; it primarily involved inserting new tasks and dependencies and updating transformation definitions and resource requirements. Figure~\ref{fig:airquality-results} illustrates a representative forecast artifact produced by the workflow after the ML-based extension.

\subsubsection{Adding new data sources from SAGE}
We further extended the workflow by adding ingestion steps for environmental measurements from the NSF SAGE sensor network~\cite{sage2021}. This required adapting data access logic and metadata handling at ingestion time while keeping the downstream analysis structure consistent. The resulting workflow demonstrates how pattern-based workflow engineering supports evolution from a single-source pipeline (OpenAQ) to a multi-source workflow (OpenAQ and SAGE) with minimal disruption.

\subsubsection{Debugging and Iterative Refinement}
Early executions of the generated workflows exposed realistic configuration and runtime errors common in distributed, containerized environments. For example, the air quality workflow initially failed during data ingestion due to missing OpenAQ API credentials inside the execution container, causing HTCondor jobs to enter a \texttt{Held} state with missing output files. Using the AI assistant, we traced the failure across Pegasus, HTCondor, and container logs, identified that environment variables were not being propagated into the container, and corrected the workflow configuration accordingly. Through this human-in-the-loop debugging process, the air quality workflow reached stable execution after seven execution attempts, illustrating how AI assistance can significantly reduce the effort required to diagnose and resolve multi-layered workflow errors.

\begin{figure*}[t]
  \centering
  \subfloat[Earthquake workflow: ingestion, analysis, aggregation.]{
    \includegraphics[width=0.46\textwidth]{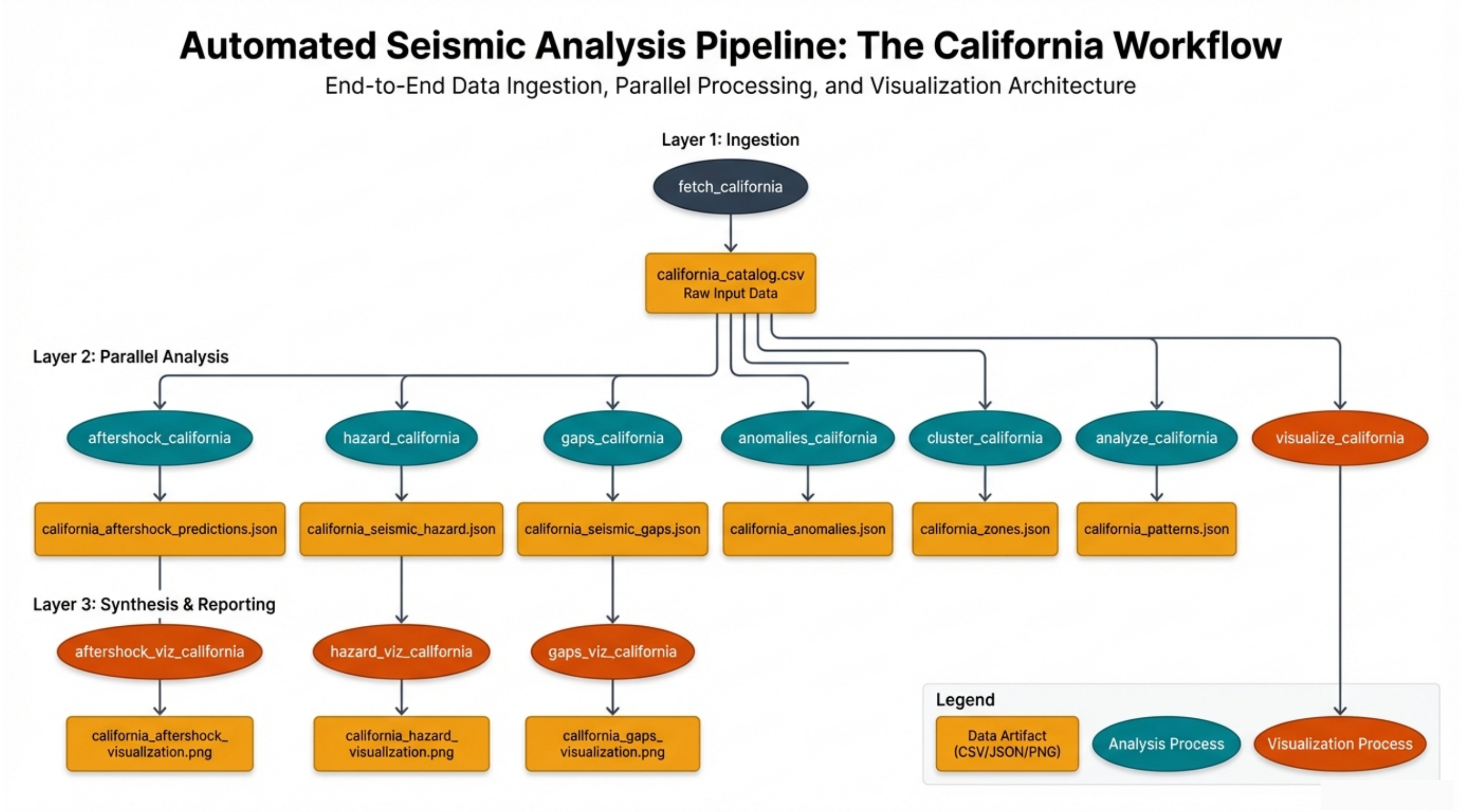}
    \label{fig:earth-dag}
  }
  \hfill
  \subfloat[Representative aftershock analysis output.]{
    \includegraphics[width=0.46\textwidth]{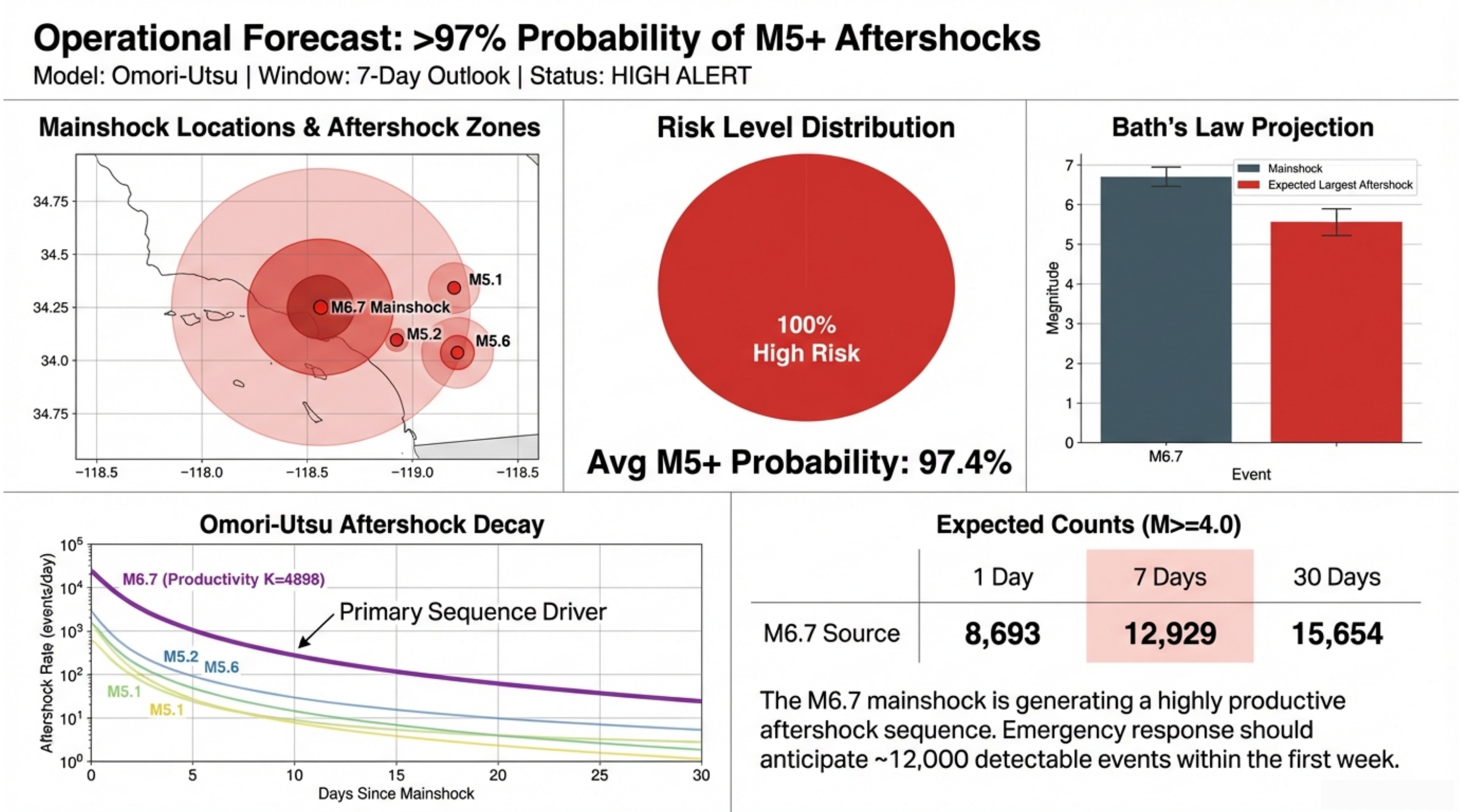}
    \label{fig:earth-aftershock}
  }
  \vspace{-5pt}
  \caption{Earthquake workflow structure and representative analytical output.}
  \label{fig:earth-workflow-results}
  \vspace{-20pt}
\end{figure*}

\begin{figure*}[t]
  \centering
  \subfloat[Soil moisture workflow: ingestion, feature joining, prediction.]{
    \includegraphics[width=0.45\textwidth]{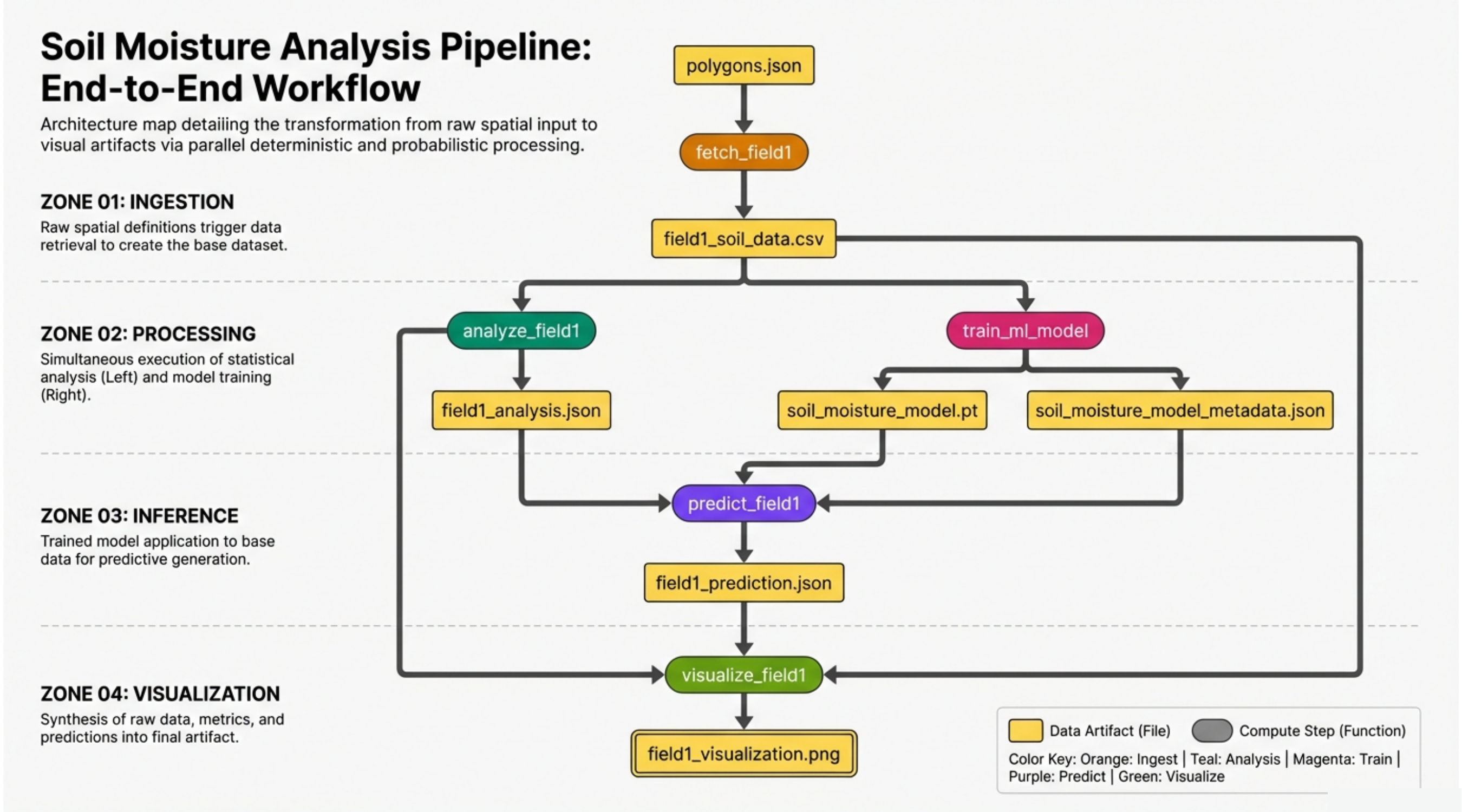}
    \label{fig:soil-dag}
  }
  \hfill
  \subfloat[Representative soil moisture prediction output.]{
    \includegraphics[width=0.45\textwidth]{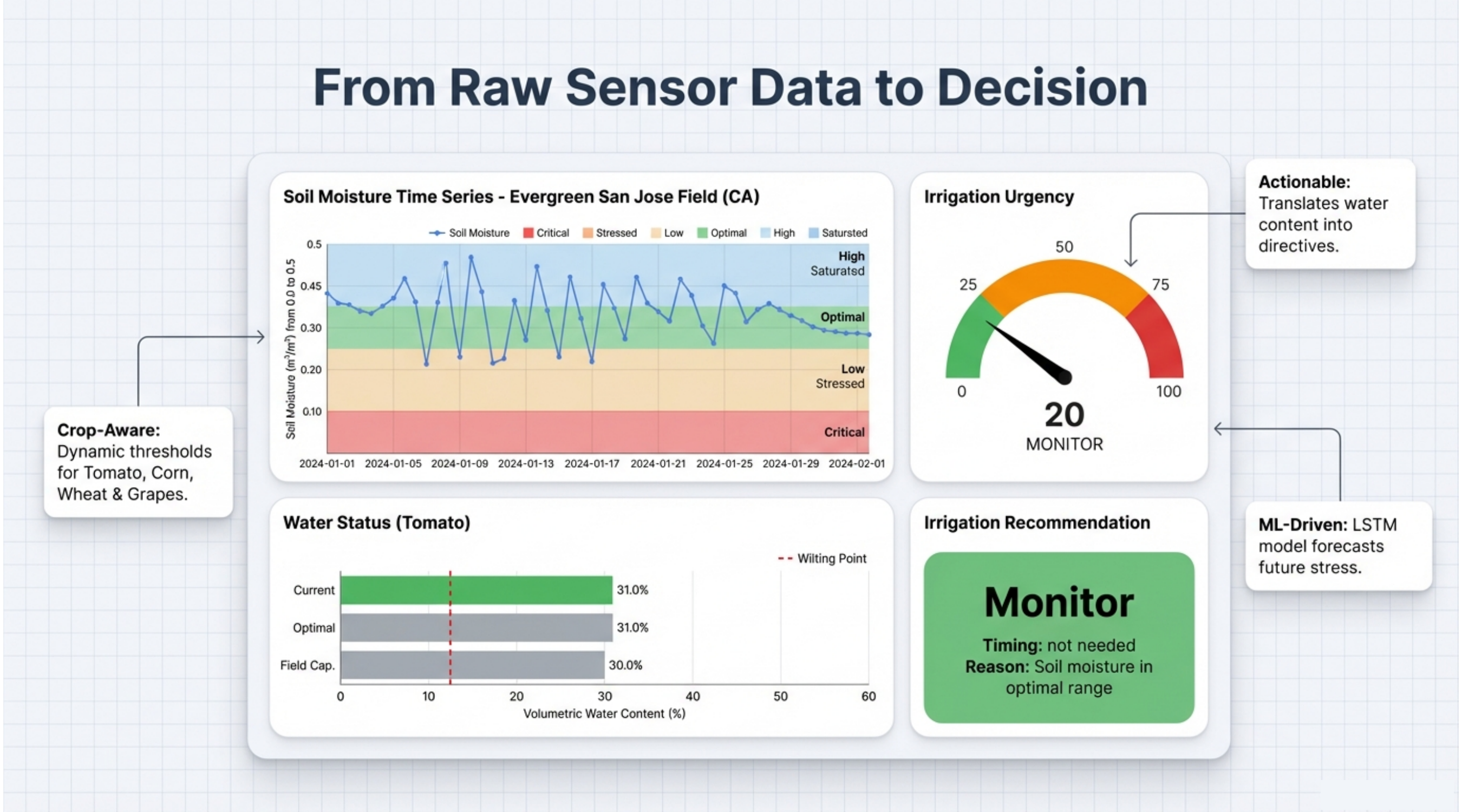}
    \label{fig:soil-output}
  }
  \vspace{-5pt}
  \caption{Soil moisture workflow structure and representative analytical output.}
  \label{fig:soil-workflow-results}
  \vspace{-20pt}
\end{figure*}

The same human-in-the-loop, AI-assisted methodology naturally extends to scaling workflows across larger task counts and execution infrastructures. As workflows grow in size, complexity, and geographic distribution, failures increasingly arise from resource heterogeneity, data movement, scheduling policies, and site-specific execution environments rather than from application logic alone. In this setting, the AI assistant can be used to iteratively reason about performance bottlenecks, resource constraints, and execution failures by correlating workflow structure with Pegasus planning output, HTCondor scheduling behavior, and site-level execution logs. Incremental scaling—first increasing task parallelism, then introducing additional execution sites or containers—allows stable workflow versions to be preserved while new scaling-related issues are isolated and resolved. This approach enables workflows to evolve from small, single-site executions to large-scale, multi-site deployments without requiring disruptive redesign, reinforcing the role of AI assistance as a tool for managing complexity in distributed scientific workflows.

\subsection{Earthquake and Soil Moisture Workflows}
\label{subsec:eq-soil}
The earthquake and soil moisture workflows were generated by reusing the already validated \textit{air quality workflow} as their immediate reference pattern, rather than reverting back to the original Orcasound workflow. This reflects a realistic workflow development process in which previously built and debugged pipelines become templates for subsequent applications.

For the earthquake workflow, the air quality pipeline was adapted by replacing the ingestion stage with USGS seismic event feeds and introducing conditional branching to trigger additional analysis for high-magnitude events. Figure~\ref{fig:earth-workflow-results} shows the earthquake workflow and a representative analytical output generated through AI-assisted, pattern-based workflow reuse.

For the soil moisture workflow, the same pipeline structure was reused, with modifications to ingest in-situ soil sensor measurements and external weather data, followed by feature joining and regression-based prediction. Figure~\ref{fig:soil-workflow-results} shows the soil moisture workflow and a representative prediction artifact produced by the workflow.

In both cases, the underlying workflow structure, execution model, and deployment configuration remained largely unchanged. AI assistance was used to translate domain-specific intent into Pegasus task definitions and to ensure that new ingestion and analysis stages were correctly integrated into the existing dependency graph.

\subsection{Integrating Edge Resources: DPUs and Raspberry Pi}
\label{subsec:edge-integration}
After validating cloud-based executions, we tailored the target workflows to incorporate edge resources. This step is central to our edge-to-core narrative: rather than designing separate edge orchestration logic, we integrated edge resources into the same Pegasus-managed workflow execution model.

\textit{1) DPU-based edge offloading on FABRIC.}
For soil moisture workflow~\cite{soilmoisture-workflow}, I/O-intensive preprocessing stages were offloaded to DPUs taking advantage of these novel edge resources. Execution placement was controlled by Pegasus site mappings, allowing tasks to be routed to DPU-enabled nodes without modifying the abstract workflow. This enabled experimentation with alternative placements in the edge to cloud continuum while preserving workflow structure and content.

\textit{2) Raspberry Pis as edge participants}
To incorporate a representative IoT edge device, we configured a Raspberry Pi to participate in the workflow data processing by connecting it to the Pegasus execution environment on FABRIC for soil moisture workflow~\cite{soilmoisture-workflow}. The Raspberry Pi performed data ingestion and lightweight preprocessing, while downstream analytics executed on cloud resources. This demonstrates that heterogeneous edge devices can become first-class participants in end-to-end Pegasus workflows using standard scheduling and execution mechanisms, again with no abstract workflow changes.

\subsection{Summary of Observations}
\label{subsec:case-study-observations}
Across the case studies, we observed consistent trends aligned with our methodology goals:
(i) pattern grounding accelerated \textit{time-to-first-workflow},
(ii) incremental refinement (baseline $\rightarrow$ integrating ML steps $\rightarrow$ adding new data sources) was easier than greenfield development,
(iii) AI assistance was most effective when adapting known patterns rather than generating workflows from scratch, and
(iv) edge integration required minimal changes, primarily in execution placement and configuration rather than workflow restructuring.
These findings motivate a user-centric evaluation focused on build/iteration effort, reproducibility, and portability rather than evaluating individual applications for peak workflow performance, which can be done by Pegasus.

\emph{This progressive reuse pattern (orcasound $\rightarrow$ air quality $\rightarrow$ earthquake and soil moisture) mirrors realistic development practices and highlights how AI-assisted workflow generation compounds productivity gains across successive applications.}

\section{Evaluation: User-Centric Productivity}
\label{sec:evaluation}

Our evaluation focuses on \textit{user productivity and iteration effort} rather than application performance. Specifically, we assess how AI-assisted, pattern-based workflow development impacts (i) time-to-first-runnable-workflow, (ii) effort required to incrementally extend workflows, and (iii) effort required to debug and stabilize generated workflows in realistic execution environments. Importantly, this evaluation reflects the experience of a \textit{novice Pegasus user}. Prior to this work, the primary exposure to Pegasus consisted of a single tutorial demonstrating how to run a basic workflow on NSF Access~\cite{ACCESS_Pegasus,access-pegasus-soil}. There was no prior experience designing multi-stage workflows, configuring HTCondor pools, or integrating heterogeneous edge resources. This context is critical for interpreting the results presented below.

\subsection{Evaluation Methodology}

We evaluate user effort using the following 
%observable 
measures:
\begin{itemize}
    \item \textbf{Prompt iterations}: the number of natural-language interactions required to generate or extend a workflow.
    \item \textbf{Execution attempts}: the number of workflow executions required to reach stable, end-to-end completion.
    \item \textbf{Wall-clock development time}: the elapsed time from initial workflow conception to stable execution.
\end{itemize}

These measures capture both cognitive effort (understanding APIs, workflow structure, and execution semantics) and practical effort (debugging configuration and runtime issues), which are central to the usability of workflow systems for sensor-driven applications. Prompt counts are reported only for the air quality workflow as a representative example; subsequent workflows exhibited similar or reduced prompting effort due to progressive reuse of validated workflow patterns.

\subsection{AI Tooling and Development Environment}

Most workflow code generation and adaptation was performed using \textbf{Claude}~\cite{anthropic_claude} (specifically Claude Opus 4.5), which was particularly effective for producing Pegasus workflow skeletons and adapting existing workflow patterns. However, due to token and usage limits encountered during extended development sessions, we also leveraged additional AI tools in practice. Specifically, \textbf{ChatGPT (Codex)~\cite{openai_codex}}(GPT-5.2) and \textbf{Gemini~\cite{google_gemini}}(Gemini 3.0 Flash) were used opportunistically, most often during execution-time debugging when interpreting Pegasus planner output, HTCondor logs, and container runtime errors. We also experimented with the open-source \textbf{Qwen3-Coder-30B}~\cite{qwen3coder2024} model and observed promising results for code generation tasks, suggesting that locally-hosted models may provide a viable alternative for teams with privacy or cost constraints. This multi-tool usage reflects a realistic development workflow in which users switch between AI assistants based on availability and task suitability, rather than relying on a single model.

All workflows were deployed, executed, and validated on the \textbf{FABRIC testbed}.
Infrastructure provisioning and configuration were performed using FABRIC notebooks and examples~\cite{fabric_pegasus_artifact}, which were also provided as contextual grounding to the AI assistant when generating code for:
(i) deploying a Pegasus cluster,
(ii) attaching and configuring FABRIC Data Processing Units (DPUs), and
(iii) connecting external edge devices, such as Raspberry Pi nodes, to the Pegasus-managed execution environment.

%\subsection{Air Quality Workflow: Quantitative Development Experience}
\subsection{Air Quality Workflow: Development Experience}

The air quality workflow serves as our primary evaluation example due to its incremental development trajectory and its role as a reference pattern for subsequent workflows.

\subsubsection{Baseline workflow construction}

Starting from the Orcasound workflow as a reference pattern, constructing the initial air quality baseline pipeline (ingest--analyze--detect anomaly) required approximately \textbf{3--4 prompts} to the AI assistant. These prompts focused on adapting data ingestion logic for OpenAQ, defining basic aggregation tasks, and generating an executable Pegasus workflow skeleton.

Although the generated workflow was syntactically correct, early executions surfaced realistic runtime and configuration issues typical of distributed, containerized environments. Resolving these issues required \textbf{seven execution attempts}, during which errors related to missing credentials, data staging, and container runtime assumptions were identified and corrected with AI-assisted debugging support.

\subsubsection{Extending to ML-based forecasting}

Once the baseline workflow reached stable execution, extending it with ML-based forecasting required an additional \textbf{3--4 prompts}. These prompts resulted in new workflow stages for feature preparation, model training or update, and prediction, while preserving the existing ingestion and preprocessing logic.

Because this extension reused an already validated workflow, the resulting pipeline stabilized more quickly, requiring only \textbf{2--3 execution attempts} to reach successful end-to-end execution. This reduction in execution iterations illustrates the benefit of incremental refinement over greenfield development.

\subsubsection{Adding SAGE data ingestion}

Incorporating SAGE sensor data into the air quality workflow required \textbf{2--3 additional prompts} to adapt ingestion logic and metadata handling. Testing and validation required \textbf{2--3 execution attempts}, primarily to verify data access and schema alignment. Notably, this extension did not require restructuring the workflow, only modifications to ingestion tasks and configuration.

\subsection{End-to-End Development Time}

Across all workflows (air quality, earthquake analysis, and soil moisture prediction), the total development effort spanned approximately \textbf{4--5 days}, corresponding to roughly \textbf{1--1.5 days per workflow}. Assuming a standard research development cadence of approximately \textbf{6--8 focused hours per day}, this corresponds to an estimated \textbf{24--40 total person-hours}, or \textbf{8--12 hours per workflow}. This time includes workflow generation, execution debugging, incremental extensions, and validation on distributed resources. 

Given the user’s novice status with Pegasus, this timeline suggests that AI-assisted, pattern-based development substantially lowers the barrier to entry for building non-trivial, multi-stage workflows. Without AI assistance, achieving comparable functionality would likely require significantly more time spent learning workflow abstractions, configuring execution environments, and diagnosing runtime failures.

\subsection{Cross-Workflow Observations}

The development experience observed for the air quality workflow was representative of the subsequent earthquake and soil moisture workflows, which reused the already validated air quality pipeline as their immediate reference pattern. Across all workflows, we observed the following trends:

\begin{itemize}
    \item \textbf{Rapid time-to-first-workflow}: Runnable workflows were generated within a small number of prompts when grounded in an existing pattern.
    \item \textbf{Decreasing iteration cost}: Later workflow extensions required fewer execution attempts due to inherited configuration and prior debugging.
    \item \textbf{Debugging dominates effort}: Most iteration effort was spent resolving execution-time configuration issues rather than correcting workflow logic.
    \item \textbf{Compounding productivity gains}: Progressive reuse (Orcasound $\rightarrow$ air quality $\rightarrow$ earthquake/soil moisture) reduced development effort for each successive workflow.
\end{itemize}

\section{Conclusions and Future Work}
\label{sec:conclusion}

This paper presented an experience-driven methodology for rapidly developing sensor-driven, edge-to-core applications by combining pattern-based workflow engineering with AI-assisted development. Using Pegasus on the FABRIC testbed, we showed how validated workflow patterns can be incrementally adapted across domains and execution environments, enabling users to construct and stabilize non-trivial sensor-driven workflows within a bounded timeframe.

Across case studies in air quality analysis, earthquake processing, soil moisture prediction, and hydrophone-based audio analytics, we demonstrated that AI assistance is most effective when grounded in proven workflow patterns and applied to iterative refinement and debugging. Our evaluation highlights how progressive reuse compounds productivity gains and lowers barriers to developing complex distributed workflows.

This work has limitations, including an evaluation focus on development effort rather than application performance and reliance on a single user’s experience. Additionally, configuring heterogeneous edge resources still requires domain expertise beyond current AI capabilities. These limitations motivate future work and exploration of tools like Kiso~\cite{Kiso2026}.

Future directions include extending the methodology to team-based and longer-lived workflows, incorporating execution feedback for automated adaptation, and integrating AI agents that proactively suggest Pegasus optimizations and placement strategies. We believe that that this approach offers a promising path toward more accessible and sustainable edge-to-core application development for the scientific community.

\bibliographystyle{IEEEtran}
\bibliography{references}

\end{document}